\documentclass[journal=jpcb,manuscript=article,layout=onecolumn]{achemso}

\usepackage{chemformula} 
\usepackage[T1]{fontenc} 
\usepackage{textgreek}
\usepackage{siunitx}
\usepackage{soul}

\author{Mario Araujo-Rocha}
\affiliation{CNRS Laboratoire de Biochimie Théorique, Institut de Biologie Physico-Chimique, Université Paris Cité, 13 rue Pierre et Marie Curie, 75005 Paris, France}
\author{Alejandro Diaz-Marquez}
\affiliation{CNRS Laboratoire de Biochimie Théorique, Institut de Biologie Physico-Chimique, Université Paris Cité, 13 rue Pierre et Marie Curie, 75005 Paris, France}
\author{Guillaume Stirnemann}
\affiliation{PASTEUR, Département de Chimie, École Normale Supérieure, PSL University, Sorbonne University, CNRS, 75005 Paris, France}
\email{guillaume.stirnemann@ens.psl.eu}

\title{On the validity of some equilibrium models for thermodiffusion}

\begin{document}
		
\begin{abstract}

When applied to binary solutions, thermal gradients lead to the generation of concentration-gradients and thus to inhomogeneous systems. While being known for more than 150 years, the molecular origins for this phenomenon are still debated, and there is no consensus on the underlying physical models or theories that could explain the amplitude of the concentration gradient in response to a given temperature gradients. Notably, there have been some attempts to relate this non-equilibrium, steady-state manifestation, to equilibrium properties of these solutions, for example, to the temperature dependence of the self-diffusion coefficient or to the solvation free-energies of each of their components. Here, we use molecular dynamics simulations on dilute solutions containing molecular size solutes, both in a thermophoretic setting as well as under equilibrium conditions, to test the validity of such models. We show that these approaches are inadequate and lead to completely uncorrelated estimates as compared to those based on the out-of-equilibrium measurements. Crucially, they fail to explain the strong mass-dependence (to which thermodynamics or single particle diffusion are insensitive) observed in the simulations and measured in the experiments. However, our results suggest an interesting correlation between the amplitude of the short-time molecular motion and that of the concentration-gradient that would deserve future investigations. 

\end{abstract}

\section{Introduction}

When subject to an inhomogeneous spatial temperature distribution, a liquid mixture's local composition usually becomes temperature (and thus spatially) dependent. This phenomenon is known as thermodiffusion, thermophoresis (especially when considering large particles in a solvent), or Soret effect\cite{Rahman2014,Kohler2016,Niether2019}. In order to explain such experimental observations, that can be quantified with a wide range of techniques\cite{Kohler2023}, a typical phenomenological approach consists of considering a particle current that is proportional to the concentration and to the temperature gradient (through a factor known as "thermal diffusion coefficient") and which adds up with the regular Fickian diffusion current, written as a being proportional (through the regular diffusion coefficient) to the concentration gradient\cite{Kohler2016}. At steady-state, the absence of net fluxes results in an exponentially distributed concentration of particles as a function of temperature, which is consistent with the experimental measurements. The temperature-dependence of the concentration is called the Soret coefficient, and it corresponds to the ratio of thermal and regular diffusion coefficients. This phenomenon is observed for many different systems, across many lengthscales, and the present contribution will exclusively discuss and focus on molecular-size diffusing objects. 

These considerations remain purely phenomenological. A vast body of experimental and computational work has attempted to go beyond this picture and characterise the molecular determinants of thermodiffusion (see, e.g. \cite{Kohler2016}). While being an "old" problem, this remains a partly open question and a timely topic, with very recent work testing phenomenological thermodynamic models using molecular dynamics simulations of Lennard-Jones particles\cite{Zimmermann2022,Gittus2023,Hoang2022}.

It is now widely accepted that the Soret coefficient can be decomposed as a sum of a purely chemical component (which depends on the interatomic interactions between the components of the mixture), a mass-dependent component, sensitive to the difference in masses between the mixture particles, and a component sensitive to the difference in the moments of inertia\cite{Kohler2016,Rutherford1987,Rutherford1989}. This decomposition was extensively studied through molecular simulations of hard-spheres and Lennard-Jones binary fluids \cite{Hoang2022}. Very recent work completed this picture by highlighting the effect of the mass dipole \cite{Gittus:2023vk} (i.e., the first moment of the mass distribution in the molecules). If we take the example of aqueous mixtures, the so-called chemical component has been related to the hydrophobic/hydrophilic nature of the solutes, in particular, to the strength of their hydrogen-bond interactions with water\cite{Niether2018,Niether2018a,Niether2019}. 

Although interesting, making such connections does not really provide an underlying molecular picture of the phenomenon. At the turn of the 20th century, Einstein's work paved the way for a molecular understanding of single particle diffusion\cite{einstein}. The diffusion coefficient was thus connected to the mean square displacement of molecules, as well as to the atomistic details of the diffusing particle (e.g. its radius) and of the diffusing medium (e.g. its viscosity). As opposed to regular diffusion, a major challenge is that thermophoresis is, in essence, a phenomenon out of thermodynamic equilibrium, which considerably limits the theoretical framework and tools that can be used. While many attempts have been made for thermodiffusion\cite{Eastman:1926ur,Kohler2016,Artola2013,Niether2019,Duhr2006a,Prigogine1952,Artola2008,Wurger2013}, there is currently no "universal", widely-accepted theoretical model that is able to quantitatively connect the Soret coefficient or the thermal diffusion coefficient to the structural and thermodynamical characteristics of the mixture, especially for molecular size solutes. We note however that models that contain at least some empirical or phenomenological ingredients were shown to correlate well with the thermophoretic behavior of hard spheres and Lennard-Jones particles\cite{Hoang2022}, although some limitations were also pointed out\cite{Zimmermann2022}. 

Despite these practical limitations, several appealing models have been proposed that rely on equilibrium considerations to explain the steady-state of a liquid system under a thermal gradient. For example, Eastman derived the following equation (written here with modern notations) for a particle immersed in a fluid\cite{Eastman:1926ur}: 
\begin{equation}
S_T = \frac{1}{k_BT}\frac{dG}{dT}
\label{eqn:dg}
\end{equation}
where $G$ is the free-energy of the particle in the solution. This idea was later reinforced by a number of experimental results from the Braun group on colloidal suspensions\cite{Duhr2006,Duhr2006a,Reichl2014}, that were shown to agree with such models, which therefore assume local equilibria along the temperature gradient. Termed differently, the particles in a solution would, therefore, tend to accumulate in the regions where the system can minimize its free-energy; that is, it follows the gradient of solvation free-energy. The validity of such a picture has often been questioned, and it was shown that its applicability could depend on the length-scale of the considered system\cite{Wurger2013}. 

Another interesting theory relates the Soret coefficient of particle 2 in a solution with 1 to the differences in the activation energy for single particle diffusion $E_a$ of the two particle types 1 and 2 in the mixture\cite{Prigogine1952}: 
\begin{equation}
S_T = \frac{E_a^2-E_a^1}{RT^2}
\label{eqn:ea}
\end{equation}

In other words, the molecules with the highest activation energies for diffusion would tend to accumulate in the cold regions (positive Soret coefficients). This equation echoes a phenomenological expression where the diffusion activation energy is replaced by the heat of transport of the species in the mixture. It is important to note that there seems to be a sign typo in the original publication\cite{Prigogine1952}, the final equation (7.9 in the original article) not being consistent with the discussion in the text. 

Measuring the single particle diffusion of molecular systems or their solvation free-energy is not straightforward in the experiments; the validity of these models has often been discussed based on some measured correlations, for example, with the solute surface area\cite{Duhr2006,Duhr2006a}, or based on theoretical arguments only\cite{Eastman:1926ur}. For example, one obvious limitation of these models is that their ingredients are mass-independent, i.e. they cannot explain, without further corrections, the mass-dependence of the Soret coefficient. In order to test these models more thoroughly, we use molecular dynamics in explicit solvent, replicating experimental thermophoretic settings and equilibrium conditions. We use two types of solutions, starting from binary mixtures of Lennard-Jones (LJ) particles as well as aqueous solutions. We focus on the dilute regime (from the perspective of the solute). 

We show that these equilibrium models are not adequate. There is no obvious correlation between the Soret coefficient and the solvation free-energy of the dilute particle or the difference between its activation energy for diffusion and that of the solvent. These models fail to reproduce qualitatively the measured trend, and their predictions are sometimes off by one order of magnitude. We finish by discussing some interesting correlations between the Soret coefficient in these mixtures, which suggest that, for very different systems ranging from the binary LJ solutions to several aqueous mixtures and for a wide range of molecular masses that can be artificially tuned in the simulation, the Soret coefficient is directly correlated by the relative difference between the amplitude of the solute motion at short timescales compared to that of the solvent.

\section{Methods}

\subsection{Systems} 

The simulations of the LJ binary mixtures considered Argon-like particles\cite{ROWLEY1975401}, with the solvent being described with $\epsilon= 0.2381$ kcal/mol, $\sigma= 3.405$~\r{A} and $m = 39.94$~u. Solute particles were identical and only differed in their $\epsilon$ values and sometimes their mass. The reference temperature was thus $T = \epsilon/k_T= 119.8$~K, with a corresponding $T^*=1$ in reduced units. As detailed below, simulations were typically performed at reduced temperatures ranging from $T^*=0.9$ to $1.1$ ($107.82$ to $131.78$~K) and reduced pressure $P^*=0.8$ (33.52 MPa). We have thoroughly checked that the system remains in a liquid phase in these conditions. 

The all-atom molecular simulations of the binary mixtures were performed at temperatures ranging from $T=300$~K to $360$~K and a pressure of 1~bar. The forcefield parameter models for trimethylamine N-oxide (TMAO)\cite{Kast2003,Holzl2016}, urea\cite{Weerasinghe2003}, methanol\cite{Gonzalez-Salgado2016}, glucose\cite{Jamali2018} and water\cite{Abascal2005}) were extracted directly from the available literature which fulfill the working temperature and pressure conditions in aqueous solutions. 

Solvent particles were randomly integrated into the simulation box through the utilization of the Packmol package \cite{Packmol}. Simultaneously, the solute particles were uniformly inserted along the z-direction of the thermal gradient, leading to an initial configuration characterized by a flat concentration profile. The Coulombic cut-off is 8.5~\AA~and the LJ cut-off is 9~\AA, with particle-particle particle-mesh (PPPM)\cite{POLAK1969} solver for the long-range electrostatic forces, in the case of dilute aqueous solutions. The particles were randomly generated using LAMMPS software version 07-Aug-19's \cite{LAMMPS} internal random atom generator for the LJ systems. The cut-off for non-bonded interactions was set to 2.5~$\sigma$ (9.534 \AA) for the LJ systems. Timesteps were 1 fs in all cases.

\subsection{Thermophoretic simulations} 
    
 All simulations under a thermal gradient were performed following a protocol described before \cite{Diaz2021}, with 12 steps from preparation to production, using the LAMMPS software version 07-Aug-19 \cite{LAMMPS} with the eHex algorithm to generate a temperature gradient\cite{Wirnsberger2015}  These include some energy minimizations, equilibration steps in successive thermodynamic ensembles, activation of a heat exchange algorithm\cite{Wirnsberger2015} and a subsequent equilibration of the temperature and concentration gradients, and finally a production stage.  
 
\begin{figure}[tbp]
 \includegraphics[width=\linewidth]{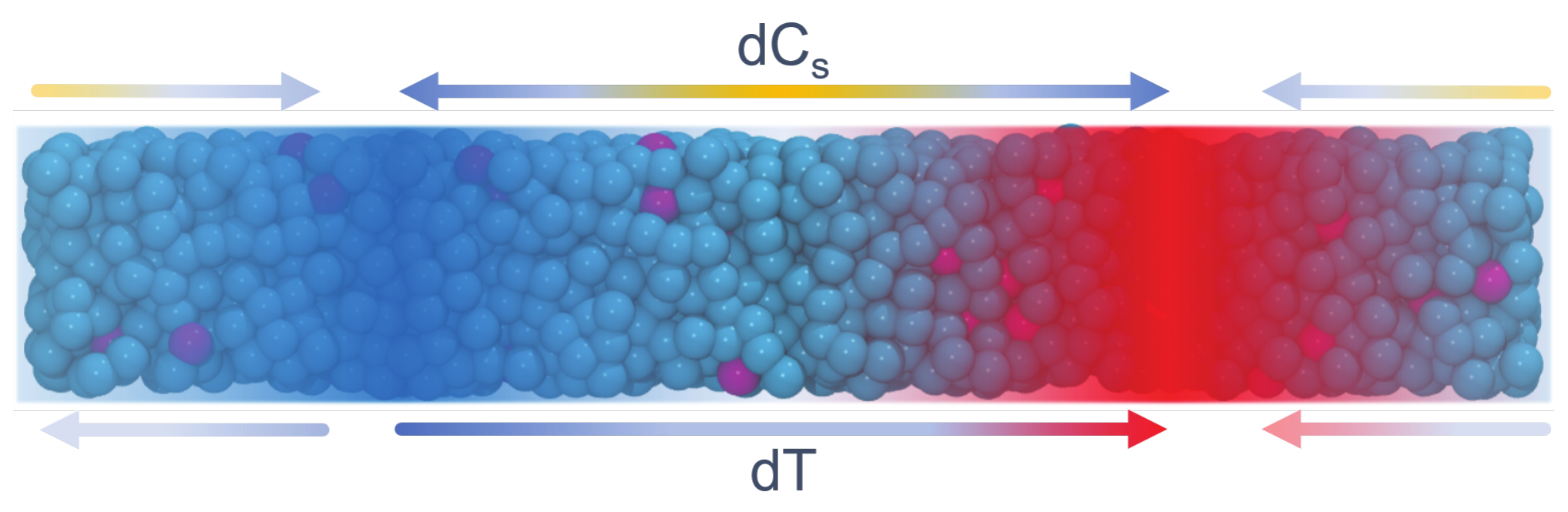}
 \caption{Schematic representation of a thermophoretic simulation (here, for LJ binary mixture). Cold and hot regions are shown in blue and red, respectively. They are positioned at one-fourth and three-fourths of the simulation box along its long $z$ axis so as to generate the temperature gradient "dT" twice, thanks to periodic boundary conditions. This in turn generates a concentration gradient "dC$_\mathrm{s}$" of solute particles (magenta) in the solvent (blue particles).}
 \label{fig:scheme}
 \end{figure}

 For the LJ mixtures, we started with boxes of $7.65~\sigma \times 7.65~\sigma \times 50~\sigma$, containing 2048 atoms of the solvent and 80 atoms of the solute. The same aforementioned protocol\cite{Diaz2021} is performed on these systems, adapting the lengths of each step to their reduced time equivalents. The thermal gradient is attained by applying the eHex algorithm of heat exchange\cite{Wirnsberger2015} in two 4 $\sigma$ long slabs over the z-axis; kinetic energy is removed from the atoms present in the region between z = 10.5 $\sigma$ and 14.5 $\sigma$ and applied to the atoms present in the region between z = 35.5 $\sigma$ and 39.5 $\sigma$. A scheme of the system and of the temperature-gradient is shown in Fig.~\ref{fig:scheme}. 

The thermophoretic simulations of the molecular binary mixtures started with boxes with $25~\AA$ length along the $x$ and $y$ axes and  $50~\AA$ along $z$. They contain 1024 molecules of water and 40 molecules of the solute. The resulting systems are identical, in terms of construction, to the LJ mixtures shown in Fig.~\ref{fig:scheme}. 
The thermostatted regions are positioned along the z-axis at 12.5 \AA~ and 27.5 \AA~ with a thickness of 4 \AA, creating two symmetrically located regions of 25 \AA.

The exchanged heat flux in the eHex algorithm was set such that the generated temperature differences between the cold and hot slab were 60~K for the aqueous solutions (Q~=~0.0375 kcal.mol$^{-1}$.fs$^{-1}$), and 0.2 (23.96~K) for the LJ systems (Q = 5 $\epsilon/\tau$, or 0.00054 kcal.mol$^{-1}$.fs$^{-1}$). 

All presented error bars for the Soret coefficient are calculated from the standard deviation of multiple simulations with different starting configurations.

\subsection{Free-energy perturbation} 

We first detail the hydration free-energy calculations for the solute molecules in dilute aqueous solutions. The simulation process involved employing cubic simulation boxes of $33.15~\AA$, filled with solvent molecules inserted randomly without spatial constraints. The system was minimized, the velocities were set to correspond to the target median temperature followed by a NPT equilibration of 200 ps at 100 kPa. The decoupling process involved progressively reducing Coulombic and LJ interactions over 40 steps using soft-core potentials by introducing a tunable parameter ($\lambda$). These simulations were performed without an induced temperature gradient.

Simulations were executed using both the LAMMPS software version 07-Aug-19 \cite{LAMMPS}, with its FEP-package, and the GROMACS package version 2019.4 \cite{GROMACS} to validate the results.  LAMMPS simulations were performed at $8.5~\AA$ Coulombic cutoff, $9~\AA$ LJ cutoff and the use of the particle-particle particle-mesh (PPPM) solver for long-range electrostatic forces. The damping parameters for the Nose-Hoover\cite{Nose1984,Hoover1985} barostat and thermostat are $P_{damp}=$~1~ps and $T_{damp}=$~0.1~ps respectively, with 3 Nose-Hoover chains and the velocity Verlet algorithm. Then, Coulombic interactions were switched off ($\lambda_{Coul}$) over 20 steps ($\delta\lambda_{Coul} =-0.05$). Once the Coulombic term was completely decoupled, the LJ potential was decoupled in the same way over the remaining windows. During this process, intermolecular non-bonded interactions were also affected, so another decoupling simulation was performed in the gas phase, and the desolvation free-energy was obtained as a difference between these two processes.

GROMACS simulations were performed at $9~\AA$ Coulombic cutoff, $9~\AA$ LJ cutoff and the particle-mesh Ewald (PME) solver\cite{Darden1993}. The damping parameters for the Parrinello-Rahman\cite{Parrinello1981} barostat and the Nose-Hoover thermostat were set at 2 ps, and the barostat compressibility were set at $4.46^{-5} bar^{-1}$, using the leap-frog integrator and 1 Nose-Hoover chain. The Coulombic and LJ interactions are decoupled over 14 steps. First, the Coulombic interactions were switched off ($\lambda_{Coul}$) over 4 steps ($\delta\lambda_{Coul} =-0.25$). Once the Coulombic term was completely decoupled, the LJ potential was decoupled, where ($\lambda_{LJ}$) decreased over 10 steps ($\delta\lambda_{LJ} =-0.1$). 

For GROMACS simulations, the Alchemical Analysis tool was employed \cite{Klimovich:2015er}, capable of handling various free energy methods, including BAR (Bennet acceptance ratio, \cite{BAR}), MBAR (Multistate Bennett acceptance ratio, \cite{MBAR}) and TI (thermodynamic integration, \cite{BOOKFRENKEL}), while LAMMPS simulations utilized a separate tool within the FEP-package \cite{FEP}, with our own adaptations made to facilitate compatibility with the Alchemical Analysis tool. 

For the LJ binary mixtures, the analyses of free-energy perturbation were done using the FEP-package on the LAMMPS software version 07-Aug-19 \cite{LAMMPS}.  As in the molecular mixtures, the system was initially minimized, then the velocities were then set to the target temperatures. This was followed by a NPT equilibration step of 1000 reduced time units (2.180 ns) with $T^*=1.0$ and $P^*=0.8$. The damping parameters for the Nose-Hoover barostat and thermostat were $P_{damp}^*$=$T_{damp}^*$=1 (2.18 ps). The LJ potential in this method is modified by a soft core to avoid singularities while annihilating atoms \cite{BEUTLER1994529}. The simulations were performed with a LJ cutoff ($r_c$) of $2.5~\sigma$ (9.534 \AA). Contrary to the method used for the molecular mixtures, there was no need of two decoupling steps, as there were no Coulombic interactions. The LJ potential was decoupled through the factor $\lambda_{LJ}$ in 20 steps ($\delta\lambda_{LJ} =-0.05$). As in the aqueous solution systems, these simulations were performed without an induced temperature gradient. In both the aqueous solution systems and the LJ systems, the presented error bars for the desolvation energies are calculated from the standard deviation of multiple simulations with different starting configurations.

%

\subsection{Mean square displacement calculations} 

A cubic box (33.15 \AA) was considered for the molecular binary mixtures. The equilibration part is replicated from thermophoretic simulations \cite{Diaz2021}, which are steps 1 to 8, in order to obtain a system in the NVE ensemble with a volume and total energy corresponding to the averages in the NPT ensemble, such that the average pressure and temperature matched the targets of the chosen thermodynamic conditions. The diffusion coefficients were determined by measuring the mean squared displacements (MSD) of the centers of mass for the solute and the solvent molecules/particles at the median temperature of 330~K. An equivalent protocol for the LJ binary mixtures was followed, with a starting simulation cubic box of size $25~\sigma$, at their median temperature of $T^*=1.0$ . As it is well-established\cite{yehhummer}, the diffusion coefficient is expected to be slightly box size-dependent because of periodic boundary effects. However, assuming that the diffusion coefficient of these molecular species can be well-described with the Debye-Stokes-Einstein model, the size-correction is temperature-independent\cite{yehhummer}, and therefore, the activation energy for diffusion is expected to be size-independent. The determination of the single-particle diffusion coefficients is typically associated with error bars on the order of 1\%, which would result in errors in the activation energy for diffusion that are much smaller than the differences observed between the solute and the solvent used in Eq.~\ref{eqn:ea}.

\section{Results and discussions}

\subsection{Determination of Soret coefficients}

We first study the thermophoretic behaviour of dilute aqueous solutions and of dilute LJ binary mixtures. In both cases, we systematically vary the solute nature and/or interaction parameters in order to simulate a range of different chemical properties for the solutes while keeping the solvent nature intact. Because these two types of solutions obviously have very different phase diagrams, we have adapted the reference temperature, density and temperature gradients to remain in the liquid phase. 

In silico thermophoretic settings are modelled as detailed in the Methods section. In short, we use a heat exchange algorithm\cite{Wirnsberger2015}  in the microcanonical ensemble, which pumps kinetic energy in one portion of the simulation box (which becomes colder than the average) and injects it in another region, which becomes the hot section of the system (Fig.~\ref{fig:scheme}). In between, a temperature-gradient is generated. The affected regions are chosen such that the gradient is unidirectional along one axis of the system (typically chosen as the $z$ axis). In general, the temperature gradient is established after a short equilibration timescale on the order of a few tens of picoseconds for the investigated systems\cite{Diaz2021}. A subsequent and much longer equilibration time\cite{Diaz2021} is then required to establish the concentration-gradient, and only then is data acquired to determine the Soret coefficient. As shown previously, these timescales correspond to those of the heat and mass transport dynamics, respectively\cite{Diaz2021}. Simulations are replicated several times (typically 10 or more) in order to determine converged values and statistical uncertainties. 

The Soret coefficient is determined after fitting the concentration profile with the following expression: 
\begin{equation}
\frac{d \ln{c_s}}{dT} = -S_T
\end{equation} 
As shown before, under the conditions employed here, the log-scale plot of the solute concentration (defined here as its molality) is typically very linear outside the thermostatted regions\cite{Diaz2021}. 

\begin{table}[htbp]
\centering
\caption{Molecular masses and Soret coefficients of molecular and LJ solutes. Data for molecular solutes is reported at the median simulation temperature of 330~K in water, while that for LJ solutes is in a binary mixture of LJ particles with solvent particles of the same size and with $\epsilon=1$ at $T^*=1.0$.}
\begin{tabular}{|l|c|c|c|}
\hline
Solute & M (g.mol$^{-1}$) & S$_T$ ($10^{-3}$ K$^{-1}$) \\
\hline
Methanol & 32.0 & 1.5 $\pm$ 0.7 \\
Urea & 60.1 & 4.4 $\pm$ 0.7 \\
TMAO  & 75.1 & 5.0 $\pm$ 0.9 \\
Glucose & 180.2 & 3.1 $\pm$ 1.1 \\
LJ Spheres - $\epsilon~=0.5$ & 39.9  & -7.6 $\pm$ 0.7 \\
LJ Spheres - $\epsilon~=1.5$ & 39.9  & 7.5 $\pm$ 2.9 \\
\hline
\end{tabular}
 \label{tab:st}
\end{table}

In Tab.~\ref{tab:st}, we show the values of the simulated Soret coefficients for several dilute aqueous solutions and binary LJ mixtures. The values are typically on the order of $10^{-3}$--$10^{-2}$~K$^{-1}$. Although Soret coefficients are most often measured for polymer or hydrocarbon mixtures mixtures\cite{AlonsoDeMezquia2014,Rutherford1987,Wittko2005,Wittko2006} or large colloidal particles\cite{Duhr2006,Duhr2006a,Reichl2014,Dhont2007}, there are a few experimental measurements for aqueous solutions, including urea\cite{Niether2018,Niether2018a} and ethanol \cite{Wiegand2007,Koniger2009,Kolodner1988,Zhang1996} (which is expected to be very close in behavior to methanol) that suggest that the Soret coefficient of the solute in these mixtures is indeed close the values found here. Similar observations can be made for LJ mixtures\cite{Reith2000a,Artola2013}. This comparison thus validates our approach and we thereby consider the simulated values as the reference that will now use to test several theoretical models.

\subsection{Solvation free-energy}

In order to determine the solvation free-energy at several different temperatures, we used standard approaches based on the progressive decoupling of the interactions between the target molecule and its environment. Simulations were performed by different simulation codes. For the aqueous solution simulations, as a natural choice, we first used the same simulation software that was employed for thermophoretic simulations, but we also complemented our study with simulations performed with Gromacs\cite{GROMACS}. For each software used, we compared three different algorithms in order to estimate the free-energy cost associated with the process: TI, BAR and MBAR. Further details on these can be found in the Methods section. 

\begin{figure}[tbp]
 \includegraphics[width=\linewidth]{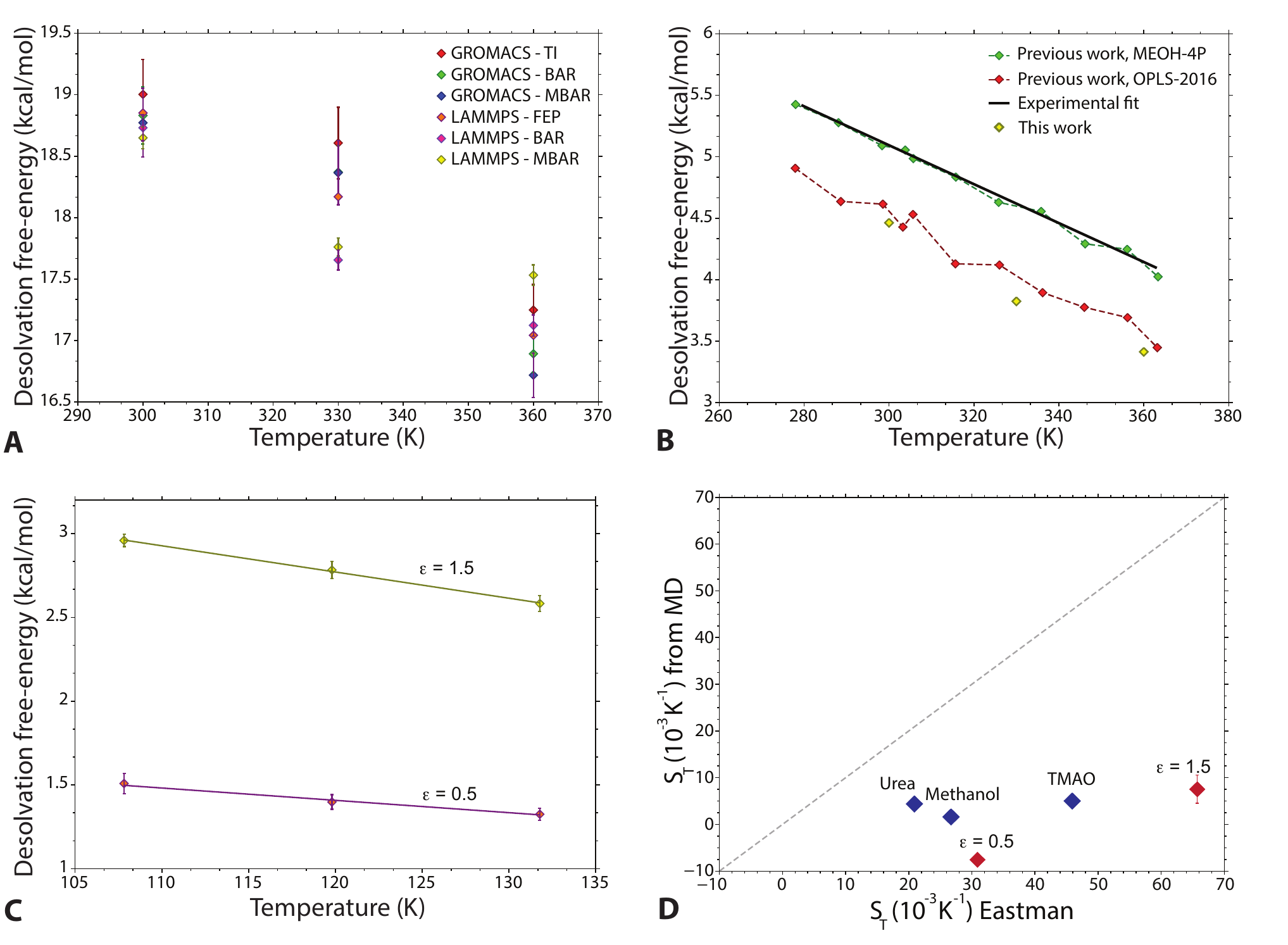}
 \caption{Solvation free-energy approach. (A) Desolvation free-energy of TMAO as a function of temperature in the dilute aqueous solution. Two different software, as well as different algorithms, are compared. (B) Desolvation energy of methanol in the dilute aqueous solution as a function of temperature, comparing results from this work (yellow) to those of previous ones (green and red, using two different forcefields) as well as the experimental estimate (black line). (C) Desolvation free-energy for the solute particles as a function of temperature for the binary LJ mixtures. (D) Comparison between the direct estimate of the Soret coefficient in the investigated dilute solutions (vertical axis) and the predictions from the free-energy model (horizontal axis).}
 \label{fig:dG}
 \end{figure}

We first take the example of a dilute aqueous solution of TMAO to compare the results from these different approaches (Fig.~\ref{fig:dG}A). Even when using the same raw data from the simulations, we notice that different algorithms give slightly different values, but the employed simulation codes give consistently similar answers. All these variations are typically within error bars, and lies within 0.4 kcal/mol from each other at 300~K and less than 1~kcal/mol at 330 and 360~K. Noticeably, they all quantitatively agree for the observed trend upon increasing temperature, with a $\approx~2$~kcal/mol increase in the solvation free-energy. 

A second important step is to verify that our simulations agree with previous simulations and experimental data when available. Because we were lacking such data for TMAO, we do this comparison for methanol, using the Laamps code now as the reference method (Fig.~\ref{fig:dG}B). Our simulations are in excellent agreement with previous simulation results using the OPLS forcefield for methanol\cite{Martinez-Jimenez:2018wf}, and are smaller in absolute value than the experimental values\cite{STAUDINGER2001561} (by about 0.5~kcal/mol). However, a very good agreement with experimental data was reported using a specifically reparametrized forcefield\cite{Martinez-Jimenez:2018wf}. Despite these small differences, a critical aspect is that the experimental temperature-dependence is very well reproduced by the previous simulations and the current ones. This provides a further validation of our approach. 

We now turn to the determination of the temperature-dependence of the solvation free-energy for a variety of solutes. In Fig.~\ref{fig:dG}C, we show the temperature dependence for LJ solute as a function of temperature. In addition to TMAO and methanol, we have also performed simulations on aqueous solutions of urea. Overall, we notice that the solvation free-energy becomes favorable as temperature increases, i.e., the solvation entropy is negative. In aqueous solutions, this is typically what is observed in the experiments\cite{STAUDINGER2001561} and in  previous simulation work\cite{Mobley:2009we, Kubo:1997to,Gallicchio:2000tn}, as there is an entropic cost to create a cavity accommodating the solute, and also a reduced entropy due to sometimes strong interactions between the water molecules and some of the solute chemical groups (for example, H-bond acceptors). 

The positive slope of the free-energy as temperature increases is thus compatible with the thermophobic nature of these solutes; however, testing Eq.~\ref{eqn:dg} reveals that the amplitude of these variations with temperature (Fig.~\ref{fig:dG}D) would lead to Soret coefficients at least one order of magnitude higher as compared to what is measured in the thermophoretic simulations. We do not observe strong variations of the results by changing the algorithm used to estimate the free-energies, and similar results are obtained for a range of solutes. In addition, we checked that performing the simulations with two different softwares gave the same qualitative answer. Therefore, such a model does not seem appropriate to explain thermodiffusion and the value of the Soret coefficient. 

\subsection{Temperature-dependence of single particle diffusion}

We then examine, for the same solutions, the temperature-dependence of the translational single particle diffusion coefficient of both the solute and the solvent particles or molecules. As done for the solvation free-energy calculations, we simulated the different systems under equilibrium conditions at several temperatures (see Methods). Diffusion coefficients were obtained from mean square displacement calculations. Their temperature-dependence was very Arrhenius-like, enabling the determination of the corresponding activation energies to test the Eq.~\ref{eqn:ea} model.

\begin{figure}[tbp]
 \includegraphics[width=0.7\linewidth]{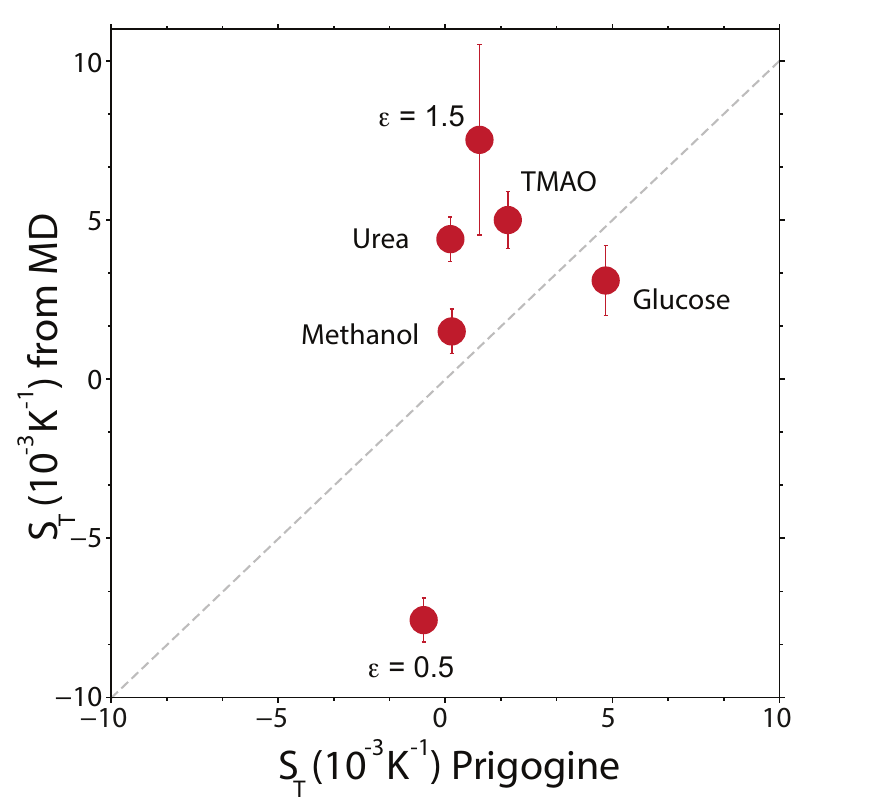}
 \caption{Activation energy for diffusion approach from Eq. ~\ref{eqn:ea}. Comparison between the direct estimate of the Soret coefficient in the investigated dilute solutions (vertical axis) and the predictions from the activation energy for diffusion model (horizontal axis).}
 \label{fig:Ea}
 \end{figure}

Starting with the aqueous solutions, we find that the activation energy for the solute diffusion is generally higher than that of water. As these typically display positive Soret coefficients, i.e., they accumulate in the cold regions, and this is consistent with the original idea of Prigogine's work that a higher activation energy for diffusion would imply that molecules would be more easily trapped in the cold regions\cite{Prigogine1952}. For the investigated solutes, the differences in the diffusion activation energy with the solvent are on the order of $RT$, which gives rise to the right orders of magnitude for $S_T$. However, as evidenced in Fig.~\ref{fig:Ea}, we fail to find any correlation between the magnitude of the difference between the activation energies for the diffusion of the solute and of the solvent and the Soret coefficient, and Eq.~\ref{eqn:ea} therefore does not appear to be valid for these solutions. Using the same methodology, we reach similar conclusions for the investigated LJ dilute mixtures (Fig.~\ref{fig:Ea}), albeit in obviously very different temperature conditions. 

As suggested before, Prigogine's model is not able to explain the effects of mass on thermophoresis\cite{Artola2008}. In particular, diffusion coefficients of a dilute particle in a given solvent are in principle mass-independent (note, however, that changing the solvent's mass could alter the solvent's viscosity and thus the diffusion coefficients of both the solute and the solvent). However, observations made from experimental and simulation measurements suggest that the mass-dependencies of thermophoresis follow phenomenological relationships in which one component of the Soret coefficient is proportional to the relative mass difference between the components of the binary mixtures\cite{Kohler2016}. The asymmetry between a change of mass in the solvent and a change of mass in the solute from the perspective of the diffusion coefficient (the first one having a small effect, the later one no effect) is a first hint that the Prigogine model would not be able to explain the phenomenological mass-dependencies. Moreover, the variations of the diffusion coefficient temperature-dependence are too small to match the large effects of mass found before. 

In order to account for the mass-dependencies, a correction to the original Progogine model has been suggested and tested on binary LJ mixtures\cite{Artola2008}. The Soret coefficient of particle 2 is a mixture with 1 was written as\cite{Artola2008}: 
\begin{equation}
S_T = \frac{E_a^2-E_a^1}{RT^2} + \frac{M^2-M^1}{M^1+M^2} \frac{E_a^2+E_a^1}{RT^2} 
\end{equation}
(we note that as Prigogine's original expression seems to be written with the wrong sign, but was taken as such when the derivation of this expression was made, we also corrected their expression when writing it here). This model was again tested for our solutions and largely overestimated the measured Soret coefficients. Indeed, for molecular solutes that are typically several times the mass of a water molecule, the mass correction simply becomes the sum of the activation energies divided by $RT^2$. However, since the activation energies for diffusion are often several times the thermal energy, the correction typically amounts to several times 1/T, whereas $S_T$ is on the order of 1/T for the dilute molecular solutes studied here. 

\begin{figure}[tbp]
 \includegraphics[width=\linewidth]{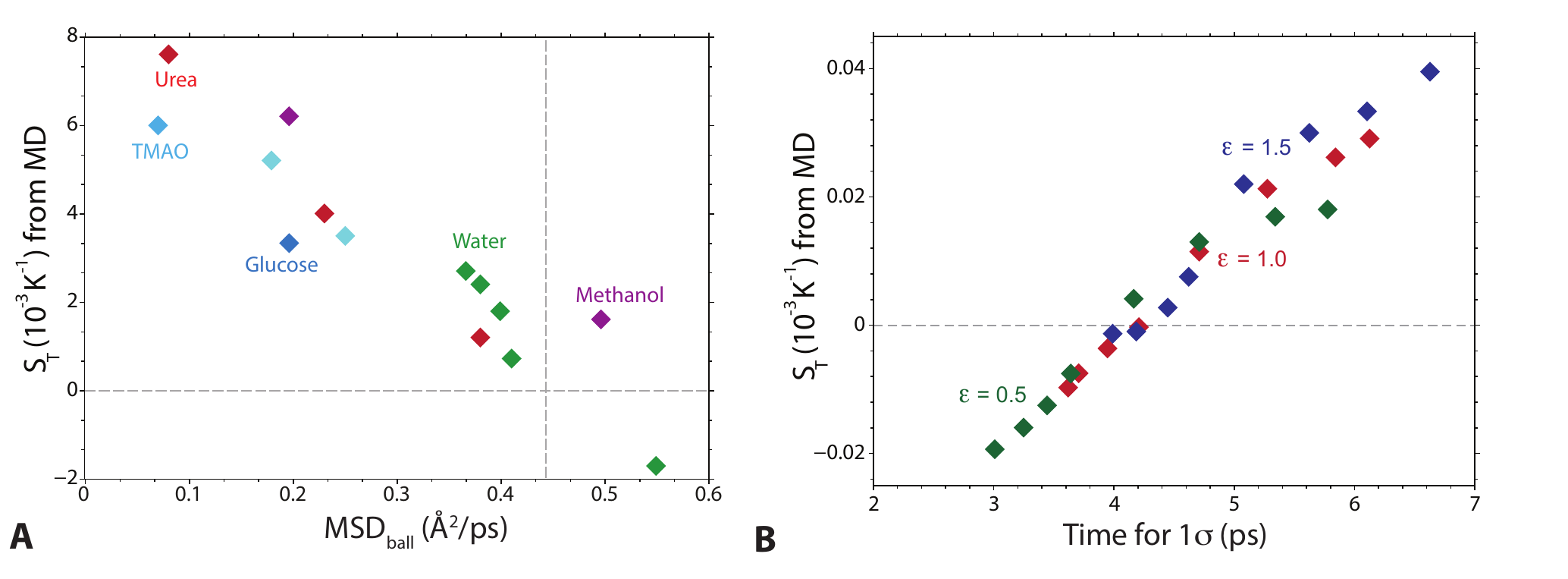}
 \caption{Correlation between the measured Soret coefficients for the investigated systems (varying both the chemical nature and the relative mass differences of the solute particles/molecules) and the short-term  motion, seen from two different perspectives. (A) Correlation  between $S_T$ at a median temperature of 330~K with the mean square displacement of the solute at 330~K after 170~fs (corresponding to a change in sign in the second derivative of the MSD, indicative of a change of regime, that was observed to be almost system independent for the dilute aqueous solutions shown here). Data for "water" solutes in water, with masses corresponding to 0.5, 1.5, 2.0, 4.0 and 8.0 times that of a regular water mass; for other solutes except glucose, where the only data point corresponds to its natural mass: mass of a water molecule, regular mass, or 4 times the regular mass. (B) Soret coefficient at a median temperature of $T^*=1.0$ vs the typical time to move by 1$\sigma$ at the same temperature. These values were obtained through the MSD measurement of simulations as described in section 2.4, tested over three values of $\epsilon$ for the solute: 0.5, 1.0 and 1.5 and eight values of solute mass: 0.33, 0.5, 0.75, 1.00, 2.00, 4.00, 6.00 and 8.00.}
 \label{fig:ball}
 \end{figure}
 
\subsection{Correlation with short-term molecular motion}

As we failed to quantitatively explain the Soret coefficients in dilute LJ mixtures and aqueous solutions with some of the previous theoretical equilibrium models, we tried to investigate possible correlations between $S_T$ and other equilibrium observables at a given temperature. Keeping in mind Prigogine's conceptual ideas, i.e., that an imbalance in the temperature-dependence of the molecular motion of molecules/particles would explain the onset of a concentration gradient, but seeking for quantities that would depend on mass, we focused on the initial regime of the molecular motion. For example, we looked at the average root mean squared displacement at a time interval corresponding to the onset of the diffusive regime after the initial ballistic region. In order to generate more data points, we (artificially) varied the solute masses and repeated the simulations for several solutes. As seen in Fig.~\ref{fig:ball}A, $S_T$ is inversely correlated to the amplitude of this motion, with molecules moving faster being associated with a lower $S_T$ value.

By taking a different perspective, we can also look at the average time it takes for a molecule to travel a certain distance. Because this picture probably makes more sense for particles of similar sizes, we performed these measurements on simulations of binary LJ mixtures for which we systematically varied the masses and the interaction energies of the solute particles. As shown in Fig.~\ref{fig:ball}B, we again find a clear correlation between $S_T$ and the timescale of a short-range initial motion covering 1~$\sigma$. Interestingly, the crossover between positive and negative $S_T$ values quantitatively agrees with the timescale of this initial motion being slower or faster as compared to that of the solvent particles. While a correlation does not imply that two quantities are directly connected to each other, we found that these observations made here for very different types of dilute solutions are encouraging and would deserve future investigations.

\section{Conclusions}

In this work, we investigate the validity of some of the models that have been proposed to explain thermophoresis based on equilibrium considerations, working here exclusively on dilute binary mixtures (where the dilute particle is considered as a solute and the concentrated one as the solvent). The first of these models is based on an idea first put forward by Eastman in the 1940s\cite{Eastman:1926ur}, and that has gained considerable attention in the last 15 years thanks to important results from the Braun group\cite{Duhr2006,Duhr2006a,Reichl2014}. Based on experimental measurements on small colloidal particles, it has been suggested that the Soret coefficient could be explained in terms of the temperature dependence of the solute solvation free-energy. The second investigated model connects the Soret coefficient to the temperature dependence of the diffusion coefficients of the solute and of the solvent molecules and was originally suggested by Prigogine\cite{Prigogine1952}, with some later refinements\cite{Artola2008}.  

 We use molecular dynamics to compute both non-equilibrium properties (such as thermodiffusion) and equilibrium ones (such as translational dynamics and solvation free-energies) and to compare them. As already shown in our own work on dilute aqueous solutions\cite{Diaz2021}, as well as in previous contributions focusing on LJ particles mixtures\cite{Reith2000a,Artola2013}, we first demonstrate how out-of-equilibrium simulations with an active heat exchange scheme can generate temperature-gradients in molecular dynamics systems, that in turn, lead to the establishment of concentration-gradients whose amplitude is compatible with experimental measurements. 

We then run and analyse molecular dynamics trajectories on the same systems but in equilibrium conditions. By using different software and algorithms, we show that we can obtain reliable and converged estimates of the solute solvation free-energies, and we examine their temperature dependence. Both for the aqueous solutions as well as for the LJ mixtures, we find that these cannot explain the Soret coefficients and that the Eastman model is not valid for the investigated molecular systems and predicts values that are one order of magnitude different from the steady-state measurements in non-equilibrium simulations. We cannot rule out that it would be correct for different systems, but we unambiguously show that this is not the case here. The same conclusions apply to Progogine's model and its variant to include mass dependencies. We find that the Soret coefficient does not correlate with the imbalance in the activation energy for diffusion between solute and solvent molecules. 

These conclusions could have perhaps  been expected. Experiments as well as simulations have evidenced the critical mass effects on the amplitude of the concentration-gradient\cite{Kohler2016}. However, free-energetic considerations, as well as single particle diffusion properties, are in principle mass-independent. A decomposition of the Soret coefficient into different terms corresponding to the intrinsic so-called chemical contribution, mass contribution, moment of inertia component, or, as recently discussed, a mass dipole contribution, is an interesting phenomenological perspective but lacks theoretical grounds; hence, a model encompassing all these effects at once is still lacking. We finish our discussion with some interesting correlations we found between the asymmetry of the short term motion of solvent and solute molecules and the amplitude of the Soret coefficient in these solutions, which we believe would deserve further investigations as they cover a wide range of chemistry and masses in very different solutions. In particular, the effect of different molecular chemistries on water dynamical properties and hydrogen-bond exchange kinetics\cite{Laage2009,Sterpone2010,Stirnemann2010a,Stirnemann2011a} could maybe be related to the thermal transport properties of the molecular solutes in aqueous solutions. 

\section{Funding}
The research leading to these results has received funding from the European Research Council under 
the European Union's  Eighth Framework Program (H2020/2014-2020)/ERC Grant Agreement No. 757111 
(G.S.). This work was also supported by the "Initiative d'Excellence" program from the French State 
(Grant "DYNAMO", ANR-11-LABX-0011-01 to GS).

\section{Acknowledgments}

The simulations presented here benefited from a local computing platform administered by G. 
Letessier, and benefited from the HPC resources of TGCC under the allocation A0070811005 made by 
GENCI (Grand Equipement National de Calcul Intensif).

\providecommand{\latin}[1]{#1}
\makeatletter
\providecommand{\doi}
  {\begingroup\let\do\@makeother\dospecials
  \catcode`\{=1 \catcode`\}=2 \doi@aux}
\providecommand{\doi@aux}[1]{\endgroup\texttt{#1}}
\makeatother
\providecommand*\mcitethebibliography{\thebibliography}
\csname @ifundefined\endcsname{endmcitethebibliography}
  {\let\endmcitethebibliography\endthebibliography}{}

\end{document}